\documentclass{article}
\begin{document}
\begin{center}
{\bf \large Fundamental particles and their interactions} \\

\medskip

by\\

\medskip

{\bf B. Ananthanarayan\\}
Centre for High Energy Physics,\\
Indian Institute of Science,\\
Bangalore 560 012, India
\end{center}

\bigskip

\begin{abstract}
In this article the current understanding of fundamental particles
and their interactions is presented for the interested non-specialist,
by adopting a semi-historical path.  A discussion on the unresolved
problems is also presented.
\end{abstract}

\vskip 2cm

\tableofcontents

\newpage
\section{Introduction}
The subject of this article is to present a sketch
of the understanding we have today of what we consider
to be fundamental particles and the interactions between them.
By adopting a semi-historical path, we shall meander through
a century or so of discoveries which have led to the establishing
of the currently accepted picture and outline some of the
outstanding unsolved problems of this picture.

At a semantic level, one is accustomed to a unique principle
or object as being fundamental.  However, in the title of
this talk, one has already invoked the plural for both
particles as well as interactions. This latter 
has been imposed on us by all that has been established
through decades of observations and experiments, and runs
somewhat contrary to the notions of philosophers who sought to
find a unique building block, {\it e. g.} Democritos of Greece
who called this the `atom', or ancient Indian philosophers who
spoke of {\it anu} in a similar spirit.  Although the name
atom has been appropriated by chemistry, and today we know
of 92 atoms associated with all the naturally occuring elements,
the idea of such an elementary or fundamental atom is an
intriguing one. It must be mentioned that the picture that will be 
presented in this discussion is one that stands in this year 2005, 
and yet has to be replaced by another picture,
if current day and future experiments reveal more elementarity.
The picture we present today
will then stand as an approximation to what is now an
unknown fundamental theory.

It is easy to list the interactions, which are the electromagnetic,
weak, strong and gravitational interactions\footnote{The
theoretical framework that describes the
first three is referred to as the `standard model'.}.
Of the fundamental
particles the only one that is familiar in day to day life is
the electron, whose presence is known from a myriad daily
experiences, as most devices that one uses implicitly uses
the properties of this fundamental particle.  In daily 
experience, this
particle interacts through the electromagnetic interactions.

This year 2005, is the `Year of Physics 2005' and commemorates three
remarkable discoveries of A. Einstein in the year 1905, namely
those of the discovery of the theory of Brownian motion,
the special theory of relativity, and the photo-electric effect.
The latter proved unequivocally that associated with the
electromagnetic interactions was a `quantum' that was to
be later called the photon.  The heuristic picture of the
electromagnetic interactions that was to emerge is one where
electrically charged particles would exchange such photons
between themselves which would give rise to the effective
force between them. Note that the idea of a `particle' associated
with light dates back to I. Newton, who spoke of a `corpuscular theory',
which was replaced by the `wave picture' in light of discoveries
of diffraction phenomena of C. Huygens.  In modern quantum theory,
the `particle-wave' duality is an inevitable feature.

In the 1905 work of Einstein is also
one of the most famous equations of all time, the energy-mass
equivalence $E=mc^2$.  It is on this equation that the
principle of nuclear energy is based, {\it i.e.}, that
in certain nuclear reactions, part of the mass of the nuclei
that participate in the reactions is released as energy.
[This is in contrast to chemical reactions, where binding energies
of molecules is liberated in a chemical reaction.]  On
the other hand, if particles are accelerated to very high energies
in a controlled manner, and are brought to a collision, in
such reactions, the energy could be converted into mass of
the daughters of the reaction process.  Einstein's theory also
predicts that for every particle, there are so-called `anti-particles',
e.g., for the electron there is a `positron' which is like an
electron in every respect, except that its electric charge is
of the opposite sign.  That this should be the case was discovered
by P. A. M. Dirac who was studying a relativistic generalization of
Schr\"odinger's equation in order to describe the motion of an
electron.  His studies led him to an equation which seemed to
admit solutions that did not permit a conventional interpretation.
Although he was himself hesitant to suggest his equation required
that the electron have a partner, the situation was clarified by
the experimental discovery by C. Anderson of a particle that was
identical to an electron, except that it had a positive electric
charge, in cosmic ray experiments.  A photon, on the other hand, is its
own anti-particle.  Therefore new kinds of
particles could be produced in this manner, subject
to certain conservation laws\footnote{For instance,
protons and neutrons belong to a family of particles
known as `baryons'.  If an electron and positron are
collided, then the reaction products must contain for every
baryon produced, an anti-baryon.}, and their interactions
studied.  This is the basis of many experiments today.  In the
past, very energetic particles from other parts of the cosmos,
so-called cosmic rays would enter the earth's atmosphere and
interact with the nuclei of the atoms therein and produce
showers of well known and new kinds of particles which could
then be detected in terrestrial or balloon borne detectors.
  
\section{The electron and the nucleus}
When one spells out the word electromagnetic interaction, it is 
manifestly a union of two other forces known through antiquity, 
{\it viz.}, the electric
and magnetic forces.  It is worth spending a little time on
the etymology of word `electric' which comes from the latin word
{\it electrum} for amber, a substance formed from wood resin,
and from which electric forces could be generated by rubbing.
Ancient humans were accustomed to electric forces generated by
the formation of clouds and the subsequent lightning bolts. 
Magnetic forces were also known by the time of the Greeks due
to the property of magnetized ore that could be used as `lodestones'
and were mined in the {\it Magnesia} prefecture of Thessaly,
one of the thirteen peripheries of Greece.  It was through the
work in recent history, in the 18th and
19th centuries, of  C. A. Coulomb, C. F. Gauss, H. C. {\O}ersted, M. Faraday, 
A. M. Amp\`ere, culminating in the work of J. C. Maxwell, that
it was shown that electricity and magnetism are manifestations
of a unified electromagnetic force, which is experienced by 
electrically charged and magnetized objects.  With the advent
of the quantum theory of M. Planck which was further developed
by N. Bohr, E. Schr\"odinger and W. Heisenberg a new theory would have
to be found to generalize `classical' electromagnetism of Maxwell.

The discovery of the electron itself as a fundamental charged particle is
attributed to the renowed English physicist J. J. Thomson
who discovered it in the year 1897, and came to be known as
the father of the electron.  He had established the 
charge to mass ratio of the electron.
Subsequently, the famous `oil-drop' experiment of
R. A. Millikan allowed one to determine the electronic charge
and it was established the mass of the 
electron\footnote{The unit chosen here is 
common in elementary particle physics: the M stands
for Mega which is equal to a million and eV stands for electron-volt,
the energy gained by an electron when it falls through a potential
of 1 Volt.  We have divided MeV by $c^2$, to express the mass in
terms of its energy equivalent via the Einstein relation.  For future
reference, another convenient unit is GeV, where G stands for
Giga which equals a billion.
We shall not use the conventional units here, which
are cumbersome for our purposes; e.g, the mass of the electron
in these units is $9.11 \times 10^{-31}$ kg.}
is 0.511 MeV/$c^2$.

It may also be noted here that
the electron carries `spin' 
\footnote{The constant $\hbar\equiv h/(2\pi)$ is 
a fundamental quantity of nature, where $h$ is called Planck's
constant.  In particular, a photon $\gamma$ of frequency $\nu$ carries
energy $E_\gamma=h\nu$.  Note that angular momentum has the
same units as energy$\times$time, which is why the spin is
associated with the same unit.}  
which is an intrinsic property of
elementary particles and has a value that is denoted by 
$\hbar/2$.
A photon on the other hand carries
a spin of $\hbar$.  Electrons are examples of fermions, named for
the Italian physicist E. Fermi since they obey `Fermi-Dirac' statistics,
while photons are examples of bosons, named for the famous
Indian physicist S. N. Bose, as they obey `Bose-Einstein' statistics.

Having measured some of
the basic properties of the electron, Thomson needed to understand
why atoms are electrically neutral under most circumstances, and
how they may be ionized.  He advanced a picture of a watermelon
like atom where a positively charged medium would have embeddeed in it,
the negatively charged electrons, much as the watermelon seeds, and
the net charges would cancel out.  This plausible picture was subjected
to experimental tests by the eminent scientist E. Rutherford.
By taking a thin gold foil and bombarding it with so-called
$\alpha-$particles arising from the radioactive decay of certain
naturally occuring elements, and by counting the scattered particles,
he was able to establish that there was an uncommonly large number
of backward scattering events of the projectiles.  The only way of
understanding this was to replace the watermelon picture of Thomson
with one where the entire positive charges were concentrated in
a very small region, later to be called a nucleus which would 
be constituted of both the `protons' as well as `neutrons', so that
an $\alpha-$ particle projectile would essentially pass unhindered
if it did not pass near the vicinity of the nucleus, or would suffer
a head-on collision and turn around and be scattered in a backward
direction.  

Thus Rutherford came to be regarded as the father of
the nucleus, and also the father of the proton, the particle associated
with the Greek word, {\it protos}, meaning first.  In other words,
Rutherford advanced the picture of the atom being consituted of
these fundamental particles, a picture which was soon revised with
the discovery by J. Chadwick of what was named as the `neutron'
which was another consitutent of nuclei.  Nevertheless, the picture
that stayed with us, until the 1950's was that only protons, neutrons
(collectively known as nucleons)
and electrons were fundamental particles.  In particular, it
was established that the masses of the proton and neutron
are 938.2 MeV/$c^2$
and 939.6 MeV/$c^2$ respectively. These also each carry spin of $\hbar/2$.

A word now about the quantum electrodynamics, the full relativistic
quantum field theory describing the interactions of electrons and 
the photon.  This is the theory which explains the quantum motion of
an electron in the field set up by another charged particle, which
is heuristically imagined as the exchange of photons between the
electron and say another electron which would be the source of
the field (so-called M{\o}ller scattering).  
One would understand the scattering of an electron in
such a field; furthermore, it would also be possible to understand
how an electron and a positron could annihilate each other by producing
energy briefly as a photon which could then produce another electron and
positron pair, in addition to the electron scattering off the field
set up by the proton.  This process was first discussed by the
Indian physicist H. J. Bhabha and is named Bhabha scattering.

In addition to the above, it is also possible that an electron
and positron could annihilate and the energy of the collision
could be converted into, say a muon and anti-muon pair, which
can be detected in laboratory detectors.  Such process
have distinct probabilities that can be computed in accordance
with the principles of quantum mechanics and special theory
of relativity (quantum field theory), and there is remarkable
agreement between theory and obesrvation.  Note here that in this
manner, particles that did not `exist' earlier can actually be
`created'.  This is a profound manifestation of energy and mass
equivalence.  In the context of quantum electrodynamics, the theory
was worked out first by Dirac, P. Jordan and others,
and put on a sound footing through the work of R. P. Feynman,
J. Schwinger and S-I. Tomonaga, and that of F. J. Dyson.

One unanswered question
was of course that which pertained to the forces between protons and
neutrons necessary to keep them inside the nucleus. In other words,
where did the inter-nucleon forces come from?  H. Yukawa proposed
that there ought to be particles, later called pions, that would 
mediate the forces between the nucleons.  These particles were
subsequently discovered in cosmic ray experiments in 1947 by
C. Powell.  These particles carry no spin.  

The question then
was whether or not these are carriers of fundamental forces just
as the photon was the carrier of the electromagnetic force,
after recognizing that this particle must have
a non-vanishing mass, unlike the photon, in order for it to
have a limited range, {\it viz.} at the inter-nucleon level\footnote{There 
are, in fact, 3 pions in nature, the charged pions
$\pi^\pm$ which weigh 139.6 MeV/$c^2$ and the neutral pion
$\pi^0$ which weighs 135.0 Mev/$c^2$.}.
The answer is no, and in the modern picture of interactions,
the internucleon force is a residue of the so-called
`strong force' that not only gives rise to the proton and
neutron in terms of constituents called quarks, while the
pions are bound states of quark-anti-quark pairs, and mediate
the effective internucleon force, as we will see in some of
the following sections.

\section{Radioactivity and the weak force}
We had already mentioned that Rutherford carried out his experiments
with $\alpha$- particles that came from radioactive decays of
some natural radioisotopes.  This decay was effectively due to
the mother nucleus spitting out a helium nucleus, which is
what the $\alpha$-particle is, and turning into a daughter nucleus.

Another kind of radioactivity was so-called $\beta$-decay, where
the particle emitted was either an electron or a positron.
A free neutron itself decays into a proton and an electron.
However, by the early 1930's it was clear that another particle
was also produced which evaded direct detection at that time, 
and was introduced by the Austrian physicist W. Pauli.  This was
named the `neutrino', the little one of the neutron by Fermi.
It had been realized at that time an unfamiliar force of nature
must be responsible for this and was named the `weak' force
which seemed to play a role only at sub-nuclear distance scales
and was so weak that its impact on day to day life on terrestrial
scales was negligible.  Indeed, in light of our prior experience,
this force would also require a mediator, and one that was
massive to ensure the short-range nature of the force, which
we today know to be the $W^\pm$ particles.  These particles are
very massive\footnote{These are about 90 times as heavy
as a proton, which in turn is about 2000 times as heavy
as an electron!}, weighing roughly
80.4 GeV/$c^2$; despite this decays can take place in accordance
with the principles of quantum mechanics and special relativity,
where a `virtual' $W$ particle can be produced, turning a neutron
into a proton and this virtual particle decaying into an electron
and anti-neutrino pair.  Here the additional energy locked up in
the larger mass of the neutron, compared to that of the proton,
is converted into the electron and anti-neutrino pair, when the
neutron gets converted into a proton.

Today, we have a complete picture of this weak interaction
and why it is short ranged, and we also know that we cannot
really speak in isolation of an electromagnetic force and
a weak force, but instead we speak of an `electro-weak' theory,
where the two forces above are different manifestations of the same.
The price to pay was the introduction of another massive cousin
of the photon, designated as the $Z^0$, which weighs 91.2 GeV/$c^2$.
This theory was sucessfully constructed by S. Glashow, A. Salam
and Weinberg (GSW) and was put on a sound mathematical footing
through the work of G. 't Hooft and M. J. G. Veltman.

The weak forces themselves are very peculiar
indeed and are unlike any other; it is the only one
that changes the particle type, {\it e.g.}
changes a neutron into a proton, then  it had been
discovered by T. D. Lee and C. N. Yang
that they violate mirror-symmetry (also known as
parity (P)) maximally, {\it e.g.}
left-handed neutrinos alone participate in the weak reactions.
Before the rise of the GSW theory siginificant work on this
was due to R. E. Marshak and E. C. G. Sudarshan and R. F. Feynman
and M. Gell-Mann, who had given the $V-A$ theory for the weak
interactions.  The weak forces alone also violate so-called
CP violation, where C stands for charge conjugation, which
related particles to anti-particles in a mathematical sense.
The weak forces are the only ones that require the mediators
of the force through massive particles.  The manner in which
these particles acquire their mass is a complicated phenomenon
and goes under the name of the `Higgs mechanism.'  Many
future experiments have as their goal, the discovery of
this particle and a detailed study of its properties.

\section{Particle zoology}
In the 1950's a whole set of new particles were identified
from cosmic ray experiments and from so-called fixed target
experiments.  It led to what was referred to as the `particle zoo',
where particles would live for a short time and decay into stable
daughters.  In particular, in cosmic ray experiments one found
showers of energetic and massive particles.  In one discovery
predating that of the pion, a particle that we now
call the muon was discovered independently by J. C. Street and
E. C. Stevenson and by C. Anderson and S. Nedermeyer.  
Today we know that it is
just like an electron except that it is about 210 times
heavier, and due to the weak interaction it decays into
an electron and into a neutrino and an anti-neutrino, 
the anti-neutrino of the electron type
and a neutrino of the muon type, in about a 
micro-second\footnote{Muons produced high up in the
atmosphere when a cosmic ray strikes a nucleus of
a nitrogen or oxygen atom (the earth's atmosphere
is mainly consituted of these elements), would not
be able to reach a terrestrial detector in their short
lifetime, but for the fact of dilatation of their life
due to their very large velocity, an effect predicted
by Einstein's theory of relativity.}.  
Much later in laboratory
experiments, a third type of `heavy electron' ($\tau$-lepton)
was discovered
by M. Perl, which is 3600 times as heavy as an electron.
Correspondingly, its lifetime compared to that of
the muon is orders of magnitude lower.  The detection of
these neutrinos themselves has itself been a rich field;
the electron type neutrinos were discovered in reactor
experiments by  F. Reines and C. L. Cowan, Jr., while
the muon type neutrinos by  L. M. Lederman, M. Schwartz
and J. Steinberger.  The existence of a $\tau$-type neutrino
has only recently been established in Fermilab experiments.

In meantime, cosmic ray experiments also showed all kinds of
`strange' events, where particles produced in pairs lived for
a relatively long time, compared to what one expected.  This
was explained by Gell-Mann and A. Pais by introducing particles
which had `strange' constituents, such that a strange and anti-strange
particle pair was produced.  Laboratory experiments also continued
to see plethora of heavier and excited particles that lived fleetingly.
This caused a serious crisis for physics which sought to see
fundamental constituents.  This was eventually resolved 
through the discovery of a theory for the strong interactions,
the subject of our next section.

\section{The strong force}
All this was put in place by assuming that these particles, and
also the nucleons were all made of underlying constituents which
we today call quarks, and the plethora was nothing but energized
bound states of these quarks.  The quarks themselves came in `flavours'
of the up, down and strange type, and carried some unknown strong charges,
and the forces were mediated by unknown carriers of the strong forces,
strong cousins of the electromagnetism's photon.  These would be called
the gluons, and the strong charges would be called `color'.  Today,
we think that all observed strongly interacting matter is the result
of the dynamics of this theory called quantum choromodynamics.  
This theory was the result of the work of a generation of scientists
primarily led by Gell-Mann.  Note
however, that the strong interaction binds quarks and gluons into
states such as the nucleons, and till date the dynamics that leads
to this is not known; this  is referred to as the `confinement' hypothesis.
Furthermore, it is due to the peculiarities of quantum field theories
that this theory behaves in such a way that when protons and neutrons
are struck by projectiles with very large energy transfer between the
projectile and the target, that the constituent quarks behave as though
they are quasi-free.  This remarkable property known as `asymptotic
freedom', was established firmly by the work of D. J. Gross, H. D. Politzer
and F. Wilczek.  Indeed, these physicists are the most recent winners
of the coveted Nobel prize for physics for the year 2004.

As time went along, it became clear that the picture needed some more
enlargement.  Indeed as there were the electron and its neutrino,
muon and its neutrino, and the $\tau$-lepton and its neutrino, it would
be necessary to group the quarks as well.  Apart from a technical detail
that we leave out here, it would be necessary to group the
up and down quarks together.  This would require that the strange
quark would require a partner; S. Glashow, J. Iliopoulos and L. Maini
would propose one called the charm quark, which was discovered in
experiments led by S. C. C. Ting and B. Richter,
and finally there would be the top and bottom quarks, discovered at
Fermilab by the CDF and D0 collaborations, and by L. Lederman respectively.  
This essentially completes the list of fundamental particles.

As mentioned earlier, the strong interactions present a very
difficult challenge in finding the solution of the microscopic
theory.  Consider for instance, the proton which is said to be
made up of 2 u quarks and 1 d quark.  It is now established that
each of these quarks weights roughly 5 MeV/$c^2$.  However, the mass
of the proton is about 938 MeV/$c^2$.  It is only through the
(as yet unsolved) dynamics of the strong interactions that
these quarks which together weigh only about 15 MeV/$c^2$ become
so much more massive.  The strange quark is believed to weigh
about 150 MeV/$c^2$, while the charm quark weighs about 1.5 GeV/$c^2$,
while the top and b quarks weigh respectively 175 GeV/$c^2$ and
4.2 GeV/$c^2$.  This incredible spread of masses of these particles
over orders of magnitude remains one of the great unsolved problems
of elementary particle physics.

\section{Synthesis}

Let us now spend a little time going over what we have learnt so far.
Elementary particles come in two varieties, those that do not participate
in the strong interactions and those that do not.  The former
are known as leptons, which participate in the weak interactions,
and of these the`charged' leptons are those that also carry electric charge
also participate in the electromagnetic interactions in a significant
manner.  On the other hand,`colored' quarks participate in the strong 
and weak interactions.  All of them carry electric charge.
All the above are spin $\hbar/2$ particles and are fermions.

As regards the forces, the electromagnetic force is carried by the photon.
The full electro-weak sector contains the $W^\pm$ and the $Z^0$
in addition.  These mediate forces between leptons and quarks.
They were experimentally discovered at CERN in 1984 for which
the Nobel prize was awarded to C. Rubbia and S. van der Meer.
The gluons are responsible for transmitting the strong force.
All the above are spin $\hbar$ particles and are bosons.
They were experimentally discovered by by the PETRA collaboration
at the German laboratory DESY.

There is one as yet undiscovered particle in the `standard model'
which is known as the Higgs.  It must weight at least 120 GeV/$c^2$.

While we have discussed in the earlier sections masses of all the particles,
we have made no mention of the masses of neutrinos.  It is notoriously
difficult to carry out measurements of the masses of these particles
and for a long time one only had upper bounds on their masses.
In the recent past, however, there is much evidence that they do indeed
possess non-vanishing masses, which comes from deep underground
experiments.  Such masses are required to resolve the `solar neutrino
problem'.  The latter, simply stated, is that fewer than
expected neutrinos reach the earth from the solar interior where
they are produced in copious numbers in reactions that power the sun.
These masses are the first concrete indication that the standard
model is incomplete.

There are still a huge number of unresolved problems.  We do not have
understanding of why there are these different types of forces.  We
do not know why their strengths are what they are.  Are they all 
seemingly different manifestations of an unknown unique force?
Even in the framework of these forces, there are unresolved questions:
for instance, why does electric charge come in multiples of a fundamental
unit?  Dirac had pointed out that if there were magnetic monopoles,
then this conclusion would have been inevitable.  However, decades
of experiments have searched for magnetic monopoles that may be
floating around in the Universe, and may have had a chance encounter
with a terrestrial detector, but have remained fruitless.
Why are there so many particles, so like each other, except for
their different masses?  Why do these elementary particles
replicate themselves?  Why are their masses what they are?

What is the role of gravitation in this picture?  How is it to be
included in a relativistic formulation?  There are many heroic
attempts, but each has its own set of technical problems and is
beyond the scope of the present discussion.  

The upcoming generation of high energy physics experiments, space
probes and the current generation of scientists are hard at work
hoping to unravel these mysteries.

\section{Acknowledgements}
This is an expanded version of talk presented at the workshop
`Vijnana Jignase (Science Discourse)' organized by the
Jawaharlal Nehru Centre for Advanced Scientific Research
and Ninasam in Heggodu, Shimoga district, Karnataka, India,
February 12-14, 2005.  It is my pleasure to thanks the organizes
K. V. Akshara, Sharath Ananthamurthy, Roddam Narasimha and
especially Srikanth Sastry for their kind invitation to present
the talk.  I thank P. N. Pandita, D. Sen and S. Vaidya for a careful reading 
of, and comments on the manuscript.

\section{Further Reading}
\begin{enumerate}
\item
G. D. Coughlan and J. E. Dodd, 
{\it The Ideas of Particle Physics : An Introduction for Scientists},
2nd edn., Cambridge University Press, 1991

\item 
Robert P. Crease and Charles C. Mann,
{\it The Second Creation: Makers of the Revolution in 
Twentieth-Century Physics}, Rutgers University Press (reprint edition),
1996

\item 
Steven Weinberg,
{\it The Discovery of Subatomic Particles},
Cambridge University Press (revised edition), 2003

\item 
B. Ananthanarayan et al., Resonance --- Journal of Science
Education, Vol. 7, No. 3, 10 (2002); 
%%CITATION = PHYS-ICS 0201043;%%
Vol. 7, No. 6, 45 (2002).
%%CITATION = PHYS-ICS 0201044;%%

\end{enumerate}

\end{document}